# Back to the Basis – Observations Support Spherically Closed Dynamic Space

**Tuomo Suntola**

*Suntola Consulting Ltd., Tampere University of Technology, Finland*

**Abstract.** A holistic view of the cosmological appearance and development of space is obtained by studying space as a spherically closed surface of a 4-sphere in a zero energy balance between motion and gravitation. Such an approach re-establishes Einstein's original view of the cosmological structure of the universe but instead of forcing space to be static with a cosmology constant, it lets it contract or expand while constantly maintaining a balance between the energies of motion and gravitation within the structure. In spherically closed dynamic space the fourth dimension, the direction of the 4-radius of the structure, is purely metric in its nature; time can be treated as a universal scalar, and the line element $cdt$ in the fourth dimension gets the meaning of the distance that space moves at velocity $c$ in time differential $dt$. The rest energy of matter appears as the energy of motion due to the motion of space in the fourth dimension, in the direction of the 4-radius of the structure. The dynamic universe approach converts Einsteinian spacetime in variable time and distance coordinates into dynamic space in absolute coordinates. All velocities in space are related to the 4-velocity of space, and the local state of rest appears as a property of the local energy system rather than as the state of an observer. Based on the zero-energy balance of whole space and the conservation of energy in interactions in space, the dynamic space approach allows the derivation of relativistic phenomena and cosmological predictions in closed mathematical form without relying on the Lorentz transformation or the relativity and equivalence principles. Mach's principle gets a quantitative expression, and the picture of cosmology is cleared: no dark energy or free parameters are needed to explain the magnitude/redshift relations of distant objects. Local systems expand in direct proportion to the expansion of whole space which results in a Euclidean appearance of distant space and explains the observed development of the surface brightness of galaxies. The rate of internal atomic processes is tied to the velocity of light which is determined by the expansion velocity and local geometry of space. All processes, like radioactive decay or buildup of large scale structures in space, have been faster in the young expanding universe. The ongoing decelerating expansion of space continues to infinity by gradually releasing the rest energy of matter. In the dynamic universe, the cycle of observable physical existence begins at cessation at infinity in the past and ends at cessation at infinity in the future.

## INTRODUCTION

Throughout the history of human thinking the origin, start and end, and the center and edges have been the most fundamental orienting questions related to our observable universe. In the Ptolemaic world the center was set to the observer's location and the edges at reasonable distances in each direction. Stellar orbits were related to the observer at the center just as they were observed.

The Copernican revolution moved the origin from the observer on the Earth to the center of the rotational system the observer belongs to. Shortly later, Kepler found the basic mathematical rules governing the motions of planets in the new coordinate system. The process was complemented by Newton who identified the laws of nature behind Kepler's equations. With the introduction of the concepts of gravitation, force and acceleration, observational astronomy had turned into physics.

The most fundamental law of physics, the conservation of energy, was implicitly included in Kepler's equations as the energy integral, the sum of the gravitational and kinetic energies. The concept of kinetic energy as "vis viva", living force, $mv^2$, was introduced by Leibniz in 1695. The conservation of the living force in elastic collisions was recognized but the fundamental nature of the conservation of energy became evident almost two centuries later through thermodynamics which also disclosed a "natural trend" towards minimum potential energy as a law of nature. Thermodynamics as well as celestial mechanics and quantum mechanics basically assume closed systems allowing the definition of the potential energy of the system in terms of the geometry and dimensions of the system. In a closed energy system the local state of rest becomes defined as a zero velocity state relative to the potential energy structure.

In present formalisms, including celestial mechanics, thermodynamics, and quantum mechanics, the velocity and the potential energy state of the whole energy system in its parent frame are assumed not to affect the energy states inside the system. Implicitly, the theory of relativity completes the idea of isolated local systems through the concepts of proper time and proper distance which, however, are primarily considered as properties of inertial observer's frame rather than properties of an energy system or potential energy frame.

In its first phase, the theory of relativity was presented as an explanation to the observed constancy of the phase velocity of light in moving frames. The extension of the special theory of relativity to the general theory of relativity created an explanation to gravity through the geometry of space-time. As a local approach based on the equivalence principle, the general theory of relativity did not give an answer to the structure of space on a cosmological scale. In 1917, as a cosmological picture of general relativity, Einstein proposed spherical space closed through the fourth dimension. Accordingly, the volume of space should be calculated as the Riemann volume of the "surface" of a four-dimensional sphere, $V = 2\pi^2 R^3$ [1]. At the cosmological level, such geometry would give a perfect view of continuous homogeneous space with no location in a special position. However, space as the surface of a 4-sphere requires the fourth dimension to be considered as a purely geometrical dimension, not a "time-like" dimension as interpreted in the theory of relativity. Furthermore, Einstein was looking for a static model. In order to remain static the 4-sphere required the famous cosmological constant to balance the gravitation of the structure. If the findings of Edvin Hubble in the 20's had been available to Einstein in 1917 his conclusions might have been different.

In his lectures on gravitation in early 1960's Richard Feynman stated:

*"...One intriguing suggestion is that the universe has a structure analogous to that of a spherical surface. If we move in any direction on such a surface, we never meet a boundary or end, yet the surface is bounded and finite. It might be that our three-dimensional space is such a thing, a tridimensional surface of a four sphere. The arrangement and distribution of galaxies in the world that we see would then be something analogous to a distribution of spots on a spherical ball."* [2]

and further

*"If now we compare this number [total gravitational energy $M_\Sigma^2 G/R$] to the total rest energy of the universe, $M_\Sigma c^2$, lo and behold, we get the amazing result that $GM_\Sigma^2/R = M_\Sigma c^2$, so that the total energy of the universe is zero. — It is exciting to think that it costs nothing to create a new particle, since we can create it at the center of the universe where it will have a negative gravitational energy equal to $M_\Sigma c^2$. — Why this should be so is one of the great mysteries—and therefore one of the important questions of physics. After all, what would be the use of studying physics if the mysteries were not the most important things to investigate".* [3]

A closer study of Feynman's "intriguing suggestion of spherically closed space" leads to dynamic space described as a spherically closed structure expanding in the direction of the radius in the fourth dimension. Such a solution shows that the rest energy of matter is the energy of motion mass has due to the expansion of space in the fourth dimension. The dynamics of space is determined by the balance of motion and gravitation in the structure, which explains the "the great mystery" of the zero-energy condition between gravitational energy and the rest energy of matter in space. Following the zero energy principle, any motion or gravitational state in space becomes related to the motion and gravitational state of whole space.

In dynamic space, contrary to the theory of relativity, time and distance can be handled as absolute coordinate quantities. Clocks in motion and clocks near mass centers run slower due to the linkage between the local energetic environment and the contribution of whole space. We need not blame time for clocks running slow close to mass centers or when moving at a high velocity in space.

Spherically closed dynamic space is studied in detail in the Dynamic Universe theory [4,5]. Predictions derived are supported by experiments equally or better than the same predictions derived from the theory of relativity and the standard cosmology model. The Dynamic Universe gives a holistic, highly ordered picture of space and the universe. The multiple forms of local expressions of energy originate from and are related to the energy built up in the contraction-expansion process of spherically closed space. Mass appears as the substance for the expression of energy — mass as such is not observable. It becomes observable through momentum when in motion and through gravitation when at finite distance from other mass.

Dynamic space has generated the rest energy of matter against a release of gravitational energy in a contraction phase from infinity in the past to singularity turning the contraction to expansion. In the expansion phase the rest energy of matter is released until it returns to zero at infinity. At infinity in the future, all motion gained from gravity in the contraction will have been returned back. Mass is conserved but it will no longer be observable because the rest energy of matter will have vanished with the cessation of motion. The energy of gravitation will also become zero owing to the infinite distances. The cycle of observable physical existence begins at cessation in emptiness and ends at cessation in emptiness.

## SPHERICALLY CLOSED DYNAMIC SPACE

### Indications of dynamic space

*The Hubble expansion*

The theory of relativity left the question of overall geometry of space unanswered and the fourth dimension was linked to time as a "time-like" dimension without direct spatial geometrical meaning. If the choice had been made ten or twenty years later, after the findings of Edwin Hubble, the result would have been different. Hubble found that light from distant objects in space are redshifted in direct proportion to their distance. He introduced the interpretation of redshift as expansion of space. When applied to the surface of a 4-sphere it implied expansion of the sphere at the velocity of light $c_4 = c$

$$v = \alpha c = \frac{D}{R_4} c_4 = \frac{c}{R_4} D = H_0 D \tag{1}$$

where $H_0$ is the Hubble constant (see FIG.1). In the standard cosmology model, the quantity $R_H = R_4 = c/H_0$ is referred to as the Hubble radius, which has the meaning of the distance, about 14 billion light years, traveled at velocity $c = c_4$ since the singularity of space.

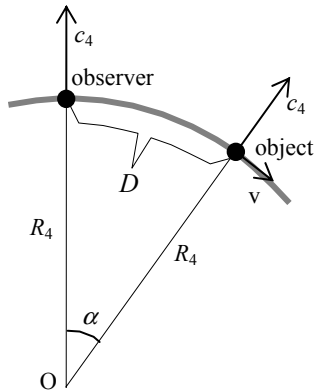

**FIGURE 1.** The Hubble law can be interpreted as a consequence of the expansion of spherically closed space at velocity $c_4$ in the direction of the radius $R_4$ in the fourth dimension. Distance *D, an arc along curved space,* is the distance between the object and the observer.

*Dynamic interpretation of the Minkowski space*

The line element in local homogeneous Minkowski space is expressed in the form
$$ds^2 = -c^2 dt^2 + dx^2 + dy^2 + dz^2 \tag{2}$$
By interpreting $c$ as a velocity in the fourth dimension, $\mathbf{c}_4$, the line element can be expressed in vector form as
$$d\mathbf{s} = i\mathbf{c}_4 dt + d\mathbf{r} \tag{3}$$
where
$$|d\mathbf{r}| = \sqrt{dx^2 + dy^2 + dz^2} \tag{4}$$
or by relating the differential $d\mathbf{r}$ in space to velocity $\mathbf{v}_r$ as
$$d\mathbf{s} = i\mathbf{c}_4 dt + \mathbf{v}_r dt = (i\mathbf{c}_4 + \mathbf{v}_r) dt \tag{5}$$

Velocity $\mathbf{c}_4$ has now the meaning of the velocity of space in the fourth dimension. In such an interpretation time is a scalar and equally applicable to velocities in space and the velocity of space in the fourth dimension.

Velocity $i\mathbf{c}_4$ *of* space in the fourth dimension gives mass $m$ at rest *in* space the momentum
$$\mathbf{p}_4 = im\mathbf{c}_4 \tag{6}$$
where the absolute value of $\mathbf{c}_4$ is equal to the velocity of light in space, or more correctly, where $\mathbf{c}_4$, as the velocity of space in the fourth dimension, determines the velocity of light in space.

When expressed like the energy of electromagnetic radiation propagating at velocity $c = c_4$, the energy due to the motion of space obtains the form
$$E = c_4 |\mathbf{p}_4| = mc_4^2 \tag{7}$$
which shows the rest energy of matter as the "energy equivalence" of momentum in the fourth dimension.

The total momentum of an object can now be expressed as the orthogonal sum of the momentum in space, in one of the three space directions, and momentum in the fourth dimension due to the motion of space as
$$\mathbf{p}_{tot} = \mathbf{p} + \mathbf{p}_4 \tag{8}$$
and, by following the concept of energy equivalence of momentum, the corresponding energy as
$$E_{tot} = c|\mathbf{p}_{tot}| = c\sqrt{|m\mathbf{c}|^2 + |\mathbf{p}|^2} \tag{9}$$
which is equal to the well known expression of the total energy introduced by the theory of special relativity through a completely different reasoning.

A thorough analysis of dynamic space shows that equation (7) applies as the fundamental definition of the energy of motion. Kinetic energy, through a change in the total momentum, describes the local work done, or potential energy lost, in obtaining a velocity in space. A general expression of the kinetic energy obtains the form
$$E_{kin} = c|\Delta\mathbf{p}| = c|m\Delta\mathbf{c} + \mathbf{c}\Delta m| \tag{10}$$

The term $m\Delta\mathbf{c}$ in equation (10) applies to kinetic energy obtained in free fall in a local gravitational frame where the velocity of local space is changed through the tilting of space, and the second term $\mathbf{c}\Delta m$ in the case of motion obtained through acceleration at a constant gravitational potential.

**Gravitation and motion of spherically closed space**

Dynamics based on a zero-energy principle shows the rest energy of matter as the energy of motion mass has due to the contraction or expansion of space in the fourth dimension, in the direction of the 4-radius. As a consequence of the conservation of energy the maximum velocity in space and the velocity light is equal to the velocity of space in the fourth dimension. The balance of the energies of motion and gravitation in spherically closed space relates the velocity of light to the gravitational constant, the total mass, and the value of the 4-radius of the structure.

*Gravitational energy in spherically closed space*

By applying Newtonian gravitational energy as the basic form gravitational energy in hypothetical homogeneous space, the total gravitational energy on mass *m* in spherically closed space is

$$E_g = -\int_{V_4} \frac{Gm\rho}{d} dV_d = -\frac{2}{\pi} \frac{GmM_\Sigma}{R_4} \int_0^\pi \frac{\sin^2 \phi}{\phi} d\phi = -\frac{GmM_\Sigma}{R_4} \cdot 0.776 = -\frac{GmM''}{R_4} \qquad (11)$$

where $G$ is the gravitational constant, $\rho$ is the average mass density in space, $dV_d$ is the volume element of space at distance $d$ from mass $m$, and $M_\Sigma$ is the total mass $M_\Sigma = \rho V_4$ in space with volume $V_4 = 2\pi^2 R_4^3$ (see FIG. 2). Mass $M''$ is located in the center of the 4-sphere and is referred to as the mass equivalence of space.

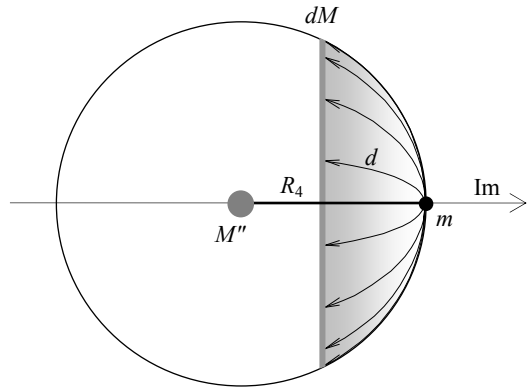

**FIGURE 2.** The gravitational energy resultant on mass *m* by mass $M_\Sigma$ distributed uniformly in the three dimensional "surface" of a 4-sphere can be calculated by integrating the gravitational energy all around the "surface". The resulting gravitational energy is equal to the gravitational energy that results from the [resulted by] mass equivalence $M''$ at distance $R''$ in the direction of the local imaginary axis. As a consequence of geometrical factors in the 4-sphere, mass equivalence $M'' = 0.776 \cdot M_\Sigma$. For mass *m* in hypothetical homogeneous space distance, $R''$ is equal to the 4-radius of space, $R_4$.

*The zero-energy balance of motion and gravitation*

Supposing the energy of motion due to the expansion of space in the direction of the 4-radius is balanced with the gravitation of the structure, an expression of the zero-energy condition can be obtained by applying equations (7) and (11) as

$$mc^2 - \frac{GmM''}{R_4} = 0 \quad \text{or} \quad M_\Sigma c^2 - \frac{GM_\Sigma M''}{R_4} = 0 \qquad (12)$$

where the first form describes the zero-energy balance of test mass *m*, and the second form the balance of the total mass $M_\Sigma$. Equation (12) relates the velocity of expansion to the gravitational constant, the total mass, and the 4-radius of space as

$$c_4 = \pm \sqrt{\frac{GM''}{R_4}} = \pm \sqrt{\frac{0.776 \cdot GM_\Sigma}{R_4}} \qquad (13)$$

where the plus and minus signs indicate that the motion can occur in both directions. In a contraction phase the motion of space is obtained against a release of gravitational energy and in an expansion phase after singularity, the motion slows down when paying back the energy of motion to gravitational energy. Space appears as a spherical pendulum is the fourth dimension. The velocity of mass due to the motion of space in the direction of the 4-radius $R_4$ does not create motion within space. The energy of motion due to velocity $c_4$ **of** space appears as the rest energy of matter **in** space.

By applying mass density $\rho$ in volume $V = 2\pi^2 R_4$ velocity $c_4$ can be expressed as

$$c_4 = \pm\sqrt{\frac{0.776 \cdot G\rho 2\pi^2 R_4^3}{R_4}} = \pm 1.246 \cdot \pi R_4 \sqrt{G\rho} = \pm 300\,000 \quad [\text{km/s}] \tag{14}$$

where the numerical value is based on $R_4 = 14$ billion light years and the density of mass in space 0.55 times the Friedmann critical mass.

When solved for time $t$ since singularity, the expansion velocity and the velocity of light in space obtain the form

$$c_0 = \frac{dR_4}{dt} = \left(\frac{2}{3} GM''\right)^{1/3} t^{-1/3} \tag{15}$$

While working against the gravitation of the structure, the velocity of expansion and, accordingly, the velocity of light in space slow down by $dc/c \approx -3.6 \cdot 10^{-11}$ /year. It can be shown that the ticking frequencies of atomic clocks and all characteristic frequencies slow down in direct proportion to the velocity of light, which makes the change in $c$ unobservable.

Time $t$ from singularity can be expressed as

$$t = \frac{2}{3} \frac{R_4}{c_0} \tag{16}$$

which means that for a Hubble radius of 14 billion light years [corresponding to Hubble constant $H_0 = 70$ [(km/s)/Mpc], the age of the expanding universe since singularity is about 9.3 billion years. In the early expanding universe, the velocity of light and the rates of all related phenomena have been higher than today. Accordingly, the predicted age, 9.3 billion years is not in a conflict with the age estimates of stars obtained with radioactive dating.

*Conservation of the total energy in interactions in space*

The picture of regular spherical space is based on a homogeneous distribution of mass assumed at the cosmological scale. Locally, mass has accumulated into mass centers. Following the zero-energy principle and the conservation of the total energy, it can be shown that accumulation of mass into mass centers in space results in the tilting [?] [bending] of the 4-surface in the fourth dimension (see FIG. 3). In tilted space the direction of the local fourth dimension (local imaginary direction) deviates from the direction of the 4-radius. As a result, the local imaginary velocity of space is reduced by factor $\cos\phi$ from the velocity of the expansion in the direction of $R_4$, the direction of imaginary axis in homogeneous space. The reduction of the local imaginary velocity of space means that the local velocity of light near mass centers is also reduced.

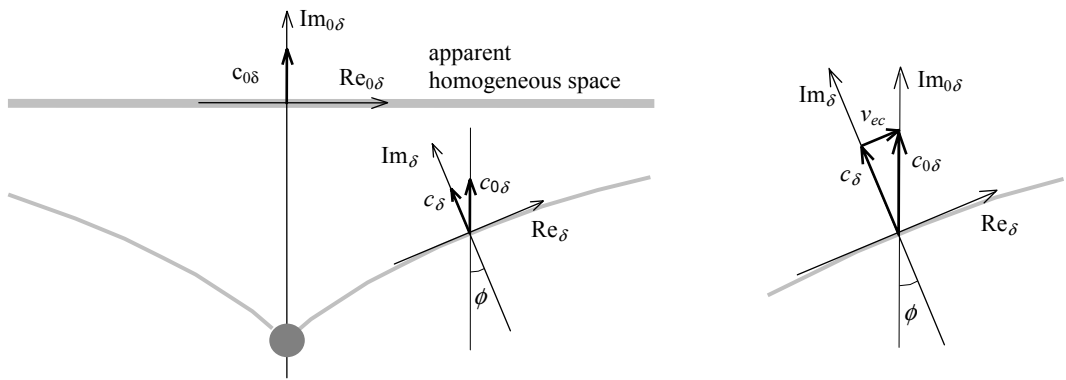

**FIGURE 3.** Space is tilted in the fourth dimension close to mass centers in space. Local complex coordinates follow the shape of space, causing the local imaginary axis, Im$_\delta$, to deviate from the direction of the imaginary axis in apparent homogeneous space, Im$_{0\delta}$. The imaginary velocity of space in the $\delta$-state, $c_\delta$, is reduced according to the formula $c_\delta = c_{0\delta} \cos\phi$, where $\phi$ is the tilting angle of space in the $\delta$-state. The orthogonal sum of the local imaginary velocity and the escape velocity, $v_{ec}$, from apparent homogeneous space is equal to the imaginary velocity of [the] apparent homogeneous space, as illustrated in the picture on the right.

By assuming the conservation of the total gravitational energy in a mass center buildup, the tilting angle of local space can be solved in terms of the distance $R$ and mass $M$ of the local center as

$$\cos\phi = 1 - \delta = 1 - \frac{MR_4}{M"R} = 1 - \frac{GM}{Rc_4^2} \tag{17}$$

where factor $\delta$ is referred to as the gravitational factor. The local velocity of light, determined by the velocity of space in the local fourth dimension, can now be expressed in terms of the gravitational factor as

$$c = c_{0\delta}\cos\phi = c_{0\delta}(1-\delta) \tag{18}$$

where $c_{0\delta}$ is the velocity of light in non-tilted space around the center.

Each gravitational center in space can be regarded as a gravitational frame where the local tilting of space is described by equation (17) relating the local tilting angle to surrounding space referred to as apparent homogeneous space of the local gravitational frame. By including the effects of all cascaded frames, the local velocity of light can be related to the imaginary velocity of hypothetical homogeneous space, $c_0 = c_4$ which is the actual expansion velocity of spherical space in the direction of the 4-radius $R_4$

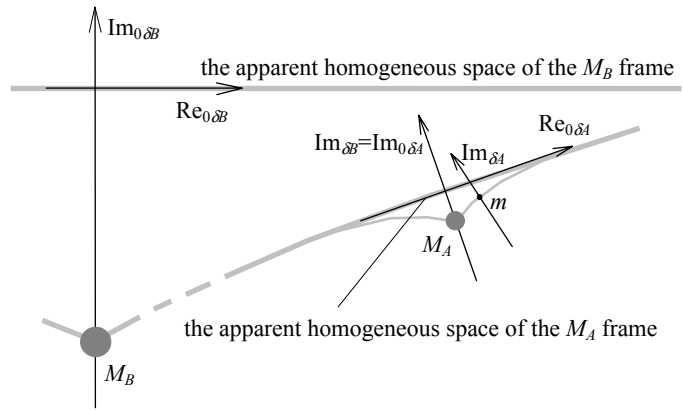

**FIGURE 4.** The apparent homogeneous space of the $M_A$-frame around mass center $M_A$ follows the direction of space in the $M_B$-frame as it would be without the $M_A$ centre.

The local velocity of light is affected by the tilting of space in the local gravitational frame and all its parent frames as

$$c = c_0(1-\delta)\prod_{n=0}^{i}(1-\delta_i) \tag{19}$$

where the gravitational factor $\delta$ is the gravitational factor in the local frame, and factors $\delta_i$ the gravitational factors of the local frame in the parent frames. (see FIG. 4).

Obviously, the reduced velocity of space in the local fourth dimension reduces the local rest momentum, the momentum of mass $m$ in the local fourth dimension. It can be shown also that a velocity in space reduces the rest momentum by reducing the internal mass, the mass "available" for the expression of energy in a state of motion. As a phenomenological explanation, the reduction of the internal mass, the mass affecting the momentum in the fourth dimension, can be understood as a consequence of the central acceleration any motion in space creates relative to the mass equivalence of space in the center of the 4-sphere. Following the conservation of energy, the internal mass $m_I$ appears as the counterpart of the effective mass $m_{eff}$ affecting the momentum of the moving object in the a space direction

$$m_I = m\sqrt{1-\beta^2} \qquad m_{eff} = m/\sqrt{1-\beta^2} \qquad m_I \cdot m_{eff} = m^2 \tag{20}$$

where $\beta = v/c$ means the velocity of the object in the local frame. The concepts of internal mass and effective mass in dynamic space are not assumptions but consequences of the conservation of energy in spherically closed space.

The internal mass *of* a moving object is the mass available for the expression of the rest energy *in* the moving object. Following the conservation of the total energy through the chain of cascaded energy frames, the motion of a local energy frame in its parent frames reduces the rest mass of an object in the local energy frame as

$$m = m_0 \prod_{n=0}^{i} \sqrt{1-\beta_i^2} \qquad (21)$$

where $m_0$ is the mass of the object at rest in hypothetical homogeneous space, and velocities $\beta_i = v_i/c_i$ are the velocities of the local frame in each of the subsequent parent frames. Combining equations (19) and (21), the rest energy of an object in a $\delta$-state in a local gravitational frame can be expressed as

$$E_{rest} = c_0 mc = m_0 c_0^2 \left(1-\delta\right) \prod_{i=0}^{n} \left(1-\delta_i\right) \sqrt{1-\beta_i^2} \quad (22)$$

which relates the rest energy of an object in a local energy frame to the rest energy of the object at rest in hypothetical homogeneous space.

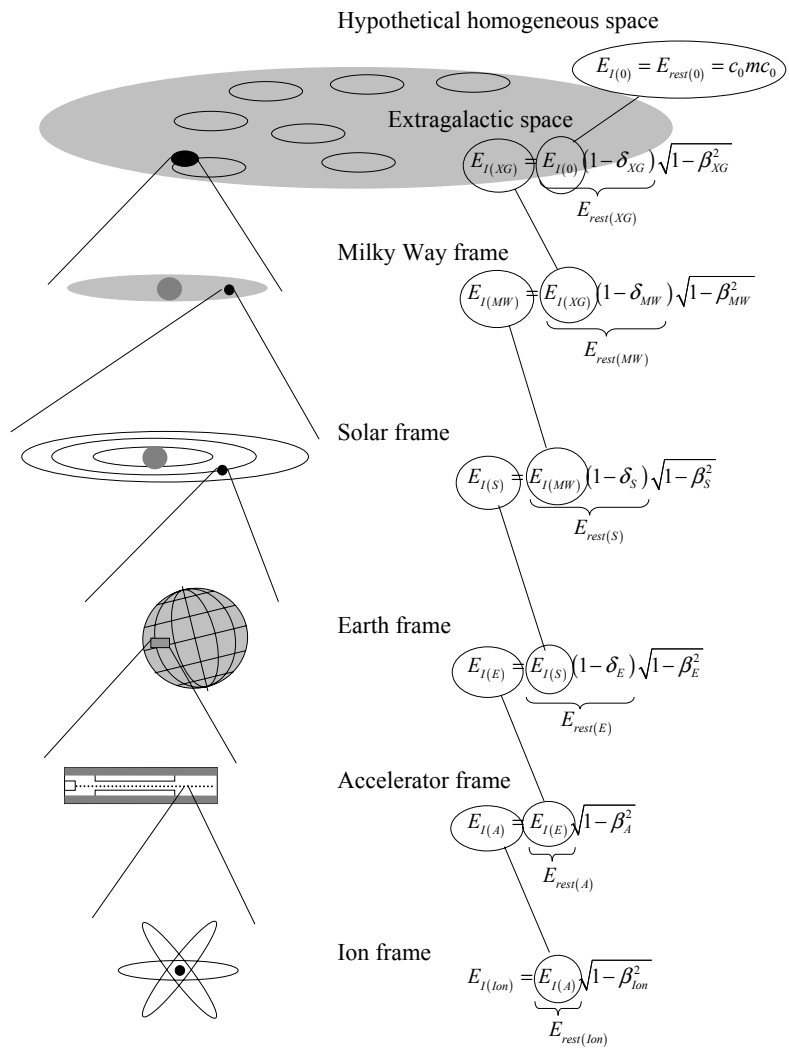

**FIGURE 5.** The rest energy of an object in a local frame is determined by the internal energy of the local frame in its parent frame. The internal energy is the imaginary component of the rest energy. The system of cascaded energy frames relates the internal energy of an object in a local frame to the rest energy of the object in hypothetical homogeneous space.

Equation (22) describes the effects of motion and gravitation on the rest energy of mass in a local energy frame in space. In order to conserve the total energy of motion due to the zero-energy expansion, any motion and local gravitational interaction in space reduce the locally available rest energy of mass.

*The rest energy of equation (22) has essentially the same effect on local physical phenomena as the concepts of proper time and proper distance in the theory of relativity. The local geometry of space given by tilting angle $\phi$ in equation (18) in each relevant gravitational frame is closely related to the space-time geometry of general relativity.*

Figure 5 illustrates the locally available rest energies in the chain of cascaded energy frames from hypothetical homogeneous space to local laboratory frames and elementary particles.

**ELECTROMAGNETIC PHENOMENA IN DYNAMIC SPACE**

*A point source as a dipole in the fourth dimension*

Space moving at $c$ in the fourth dimension allows the study of a point source as a dipole in the fourth dimension. When solved for the energy of one cycle of radiation emitted by a dipole, the standard solution of Maxwell's equations can be written in form

$$E_\lambda = \frac{P}{f} = \frac{N^2 e^2 z_0^2 \mu_0 16\pi^4 f^4}{12\pi c f} = N^2 \left(\frac{z_0}{\lambda}\right)^2 \frac{2}{3}\left(2\pi^3 e^2 \mu_0 c\right) f \quad (23)$$

where $N$ is the number of unit charges $e$ oscillating the dipole, $z_0$ is the length of the dipole, $c$ is the velocity light, and $f$ is the frequency of oscillation of the emitted radiation. Equation (23) relates the length of the dipole to the wavelength emitted and applies the vacuum permeability $\mu_0$ instead of the vacuum permittivity $\varepsilon_0$. Equation (23) assumes sinusoidal oscillation of the charge in the dipole, the factor 2/3 is the ratio of average power density to the power density in the normal plane of the dipole. Equation (23) shows that the energy of one cycle of radiation from a dipole of a fixed length is directly proportional to the frequency of radiation emitted [6]

$$E_\lambda = F\left(N, z_0, D_P\right) \cdot \left(2\pi^3 e^2 \mu_0 c\right) f \quad (24)$$

where $F(N,z_0,D_P)$ is a factor characteristic to the dipole and the input current. Factor $2\pi^3 e^2 \mu_0 c$ in equation (24) has the dimensions of Planck's constant and a numerical value close to Planck's constant

$$h' = 2\pi^3 e^2 \mu_0 c = 5.997 \cdot 10^{-34} \, [\text{kgm}^2/\text{s}] = h/1.1049 \quad (25)$$

In dynamic space moving at velocity $c$ in the fourth dimension, any point source moves one wavelength in a cycle in the direction of the fourth dimension. Accordingly, a point source at rest in space can be considered as a dipole with the length of one wavelength in the fourth dimension. According to equation (23) one electron transition ($N$=1) in a point source ($z_0/\lambda = 1$) emits a cycle of electromagnetic radiation with energy

$$E_{\lambda(0)} = F\left(D_P\right) \cdot h' \cdot f \quad (26)$$

Because all space directions are perpendicular to a dipole in fourth dimension, the power distribution factor $D_P$=1. An electron transition in an atom is a quantum jump rather than a sinusoidal oscillation, which means that we shall assume an energy transition factor in equation (26). A comparison of equations (24) and (26) to Planck's equation suggests an energy transition factor of 1.1049 which allows the expression of equation (26) in form

$$E_\lambda = 1.1049 \cdot h' \cdot f = h \cdot f = 1.1049 \cdot 2\pi^3 e^2 \mu_0 c \cdot f = h_0 c \cdot f = \frac{h_0}{\lambda} \cdot c^2 = m_\lambda \cdot c^2 \quad (27)$$

where

$$h_0 = 1.1049 \cdot 2\pi^3 e^2 \mu_0 = \frac{h}{c} \quad (28)$$

is referred to as the intrinsic Planck's constant with dimensions [kg·m], i.e. the factor $m_\lambda = h_0/\lambda$ in the last form of equation (27) has the dimensions of mass [kg]. The factor $h_0/\lambda$ appears as the mass equivalence of radiation which shows the energy of a quantum of radiation in the same form as the rest energy of matter. Application of equation (28) in the definition of the fine structure constant shows the fine structure constant as a purely geometrical factor without connection to any physical constant

$$\alpha \equiv \frac{e^2}{2h\varepsilon_0 c} = \frac{e^2 \mu_0}{2 \cdot 1.1049 \cdot 2\pi^3 e^2 \mu_0} = \frac{1}{2 \cdot 1.1049 \cdot 2\pi^3} \approx \frac{1}{137} \quad (29)$$

The finding, that the traditional Planck's constant is proportional to the velocity of light is of special importance in dynamic space because $c$ is not constant but varies with time and location in space.

*The unified expression of energy*

Equation (27) gives the energy of a quantum of radiation in the form $E_\lambda = m_\lambda \cdot c^2$, the same form as that of the rest energy of matter. The meaning of a quantum of radiation appears as the energy emitted by a single transition of a unit charge in a point source in dynamic space.

By applying the vacuum permeability $\mu_0$ instead of vacuum permittivity $\varepsilon_0$ and the fine structure constant $\alpha$ in equation (29), the expression of the Coulomb energy obtains the forms

$$E_{EM} = \frac{q^2 \mu_0}{4\pi r} c^2 = N^2 \alpha \frac{h_0}{2\pi r} c^2 = m_{EM} c^2 \quad (30)$$

where $m_{EM} = e^2\mu_0/4\pi r$ is referred to as the mass equivalence of electromagnetic energy with dimensions of [kg]. Equations (27) and (30) do not take into account the effect of the system of cascaded energy frames and the related effects of the local velocity of light. When these are included, equations (27) and (30) obtain the form equal to equation (22) which results in unified expressions of energy

The rest energy of mass:      $E_{rest} = c_0 m c$      (31)

The Coulomb energy:      $E_{EM} = N^2 \alpha \frac{h_0}{2\pi r} c_0 c = c_0 m_{EM} c$      (32)

The energy of a quantum :      $E_\lambda = \frac{h_0}{\lambda} c_0 c = c_0 m_\lambda c$      (33)

Kinetic energy of mass:      $E_{kin} = c_0 |\Delta m \mathbf{c}|$      (34)

In equations (31)…(34), $c_0$ accounts for the expansion velocity of space, $c$ the effect of the local velocity of light, and $m$, the mass or mass equivalence, the substance for the expression of energy. The local velocity of light $c$ on the Earth can be estimated to differ by about one ppm from the velocity of light in hypothetical homogeneous space, $c_0$. We may also note, for example [e.g.], that a release of Coulomb energy in an accelerator, where $c$ is constant, releases part of the mass equivalence of the electromagnetic energy through a change in $r$ in equation (32). The mass equivalence released is exactly the increase of mass, the buildup of effective mass, in the kinetic energy in equation (34) [equal to equation (10)].

*The hydrogen atom and the characteristic emission frequencies*

By applying the intrinsic Planck constant defined in equation (28), the standard non-relativistic expression of energy states of electrons in a hydrogen-like atom solved from Schrödinger's equation can be expressed in the form

$$E_{Z,n} = \frac{e^4 \mu_0^2}{8 h_0^2} \left(\frac{Z}{n}\right)^2 m_e c c_0 = \frac{\alpha^2}{2} \left(\frac{Z}{n}\right)^2 E_e \tag{35}$$

where $E_e$ is the rest energy of an electron in the nucleus energy frame. With reference to equation (22), the rest energy of electron in an atom is a function of motion and gravitational state of the atom. By applying equation (22), Balmer's formula for the characteristic frequencies of hydrogen like-atoms obtains the form

$$f_{(n1,n2)} = \frac{\Delta E_{(n1,n2)}}{h_0 c_0} = Z^2 \left[\frac{1}{n_1^2} - \frac{1}{n_2^2}\right] \frac{\alpha^2}{2 h_0} m_e c = f_{0(n1,n2)} \prod_{i=1}^{n} (1-\delta_i)\sqrt{1-\beta_i^2} \tag{36}$$

where factors $\delta_i$ and $\beta_i$ define the state of gravitation and motion of the atom, and

$$f_{0(n1,n2)} = Z^2 \left[\frac{1}{n_1^2} - \frac{1}{n_2^2}\right] \frac{\alpha^2}{2 h_0} m_{e(0)} c_0 \tag{37}$$

where $c_0$ is the expansion velocity of space, and $m_{e(0)}$ is the mass of an electron at rest in hypothetical homogeneous space. As shown by equations (36) and (37), the characteristic frequency of a specific transition in an atom is a function of both the motion and gravitation of the atom.

*Equation (36) shows the relativistic effects of motion and gravitation as an integral part of the quantum mechanical solution of the characteristic frequencies. It also combines the coordinate time scales in different frames like the Earth Centered Inertial Frame applied in satellite systems and the Solar Barycenter Frame applied in observations in the solar system and extends the coordinate time structure from intergalactic scales to laboratory frames on the rotating Earth or anywhere in space.*

Because the expansion velocity of space is subject to a gradual decrease with the expansion of space, the reference frequency $f_{0(n1,n2)}$ in equation (37) declines with time as

$$f_{0(n1,n2)} = Z^2 \left[\frac{1}{n_1^2} - \frac{1}{n_2^2}\right] \frac{\alpha^2 m_{e(0)}}{2 h_0} \cdot \left(\frac{2GM''}{3}\right)^{1/3} t^{-1/3} \tag{38}$$

where $G$ is the gravitational constant and $M'' = 0.776 \cdot M_\Sigma$ is the mass equivalence of spherically closed space. As shown by equation (36), the characteristic frequency is directly proportional to the local velocity of light, $c$, which means that in local measurements based on atomic clocks, the velocity of light is observed as constant.

When solved for the characteristic wavelength, Balmer's formula obtains the form

$$\lambda_{(n1,n2)} = \frac{c}{f_{(n1,n2)}} = \frac{\lambda_{0(n1,n2)}}{\prod_{i=1}^{n} \sqrt{1-\beta_i^2}} \tag{39}$$

where

$$\lambda_{0(n1,n2)} = \frac{c_0}{f_{0(n1,n2)}} \tag{40}$$

Accordingly, the characteristic wavelength is subject to an increase due to the motion of the emitting atom, but it is not affected by the expansion of space or the local gravitational state (or the velocity of light). The characteristic wavelength is directly proportional to the Bohr radius, which obtains the form

$$a = \frac{h_0}{\pi \mu_0 e^2 m_e} \tag{41}$$

which shows that like the emission wavelength the atomic radius is also independent of the expansion of space.

*The reduction of the frequencies of atomic clocks in motion and near mass centers in space are consequences of the energetic state, the state of gravitation and motion (the velocity) of the clock. The velocity of the clock means velocity relative to the state of rest in the energy frame the clock is moving — not the velocity relative to an observer as taught by the relativity theory. The frequency of a clock is not affected by acceleration, only the velocity and the gravitational potential apply. There is no place for the Lorentz transformation or the equivalence principle in the Dynamic Universe.*

*All energy states in space are related to the reference at rest in hypothetical homogeneous space. The unified expressions of energy apply in all local frames in space and manifest the zero energy principle as a universal law of nature. There is no place for the principle of relativity in the Dynamic Universe.*

Gravitational red and blue shift are real changes in the frequency of the characteristic emission. According to equation (36), the frequency of a radiation emitting source moving at velocity $\beta$ at gravitational state $\delta$ in a local gravitational frame is

$$f_{\delta,\beta(DU)} = f_{0,0}(1-\delta)\sqrt{1-\beta^2} \approx f_{0,0}\left(1-\delta-\frac{1}{2}\beta^2-\frac{1}{8}\beta^4+\frac{1}{2}\delta\beta^2\right) \qquad (42)$$

The corresponding equation derived from the general theory of relativity is

$$f_{\delta,\beta(GR)} = f_{0,0}\sqrt{1-2\delta-\beta^2} \approx f_{0,0}\left(1-\delta-\frac{1}{2}\beta^2-\frac{1}{8}\beta^4-\frac{1}{2}\delta\beta^2-\frac{1}{2}\delta^2\right) \qquad (43)$$

In the Earth gravitational frame (the ECI frame), on the Earth geoid, as well as on an Earth-orbiting satellite, the relative difference between the frequencies given by equations (42) and (43) is smaller than $10^{-18}$.

The frequency of electromagnetic radiation is not changed when propagating to a different gravitational potential. The wavelength of the received radiation is observed changed due to a different velocity of light at the receiver at a different gravitational potential.

At cosmological distances, the wavelength of radiation propagating in space is subject to increase in direct proportion to the expansion. Since the emission wavelength of atoms in expanding space is not subject to increase, the characteristic radiation from distant objects is observed redshifted relative to the reference emission at the location of the observer.

## COSMOLOGICAL APPEARANCE OF SPHERICALLY CLOSED SPACE

### *Optical distance and the Hubble law*

As a consequence of the identical velocities of space along the 4-radius and electromagnetic radiation in space, the optical distance $D$, which is the distance traveled by light from object $A_1$ to object $B$ in space (in the tangential direction), is equal to the corresponding change of the radius (see FIG. 6)

$$D = R_{4(2)} - R_{4(1)} \qquad (44)$$

The increase of the wavelength of radiation propagating in space is assumed to increase in direct proportion to the increase of distances in space, which defines the redshift as

$$z = \frac{\lambda - \lambda_1}{\lambda_1} = \frac{R_{4(2)} - R_{4(1)}}{R_{4(1)}} = \frac{D}{R_{4(1)}} = \frac{D}{R_{4(2)}}\frac{R_{4(2)}}{R_{4(1)}} \qquad (45)$$

The differential propagation distance of radiation in space can be expressed in terms of the distance angle $\alpha$ (see FIG. 6) as

$$ds = c \cdot dt = R_4 d\alpha \qquad (46)$$

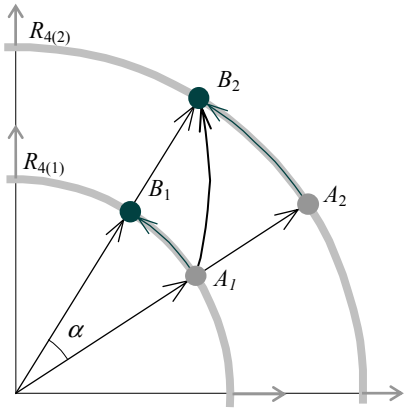

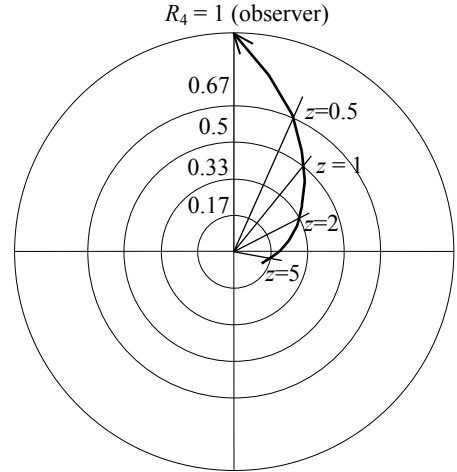

**FIGURE 6.** Propagation of electromagnetic radiation from object $A_1$ at $R_{4(1)}$ to object $B$ at $R_{4(2)}$.

**FIGURE 7.** The length of the $R_4$ radius and the location of objects for redshifts $z = 0$ to 5. Location $R_4 = 1$ is the observer's location. The optical distance to the object is the tangential length of the path, which is equal to the difference between the present $R_4$ radius and the $R_4$ radius as it was when the light was emitted. For example, the light emitted at $R_4 = 0.17$ in the drawing is redshifted by $z = 5$ after traveling the distance $D = (1-0.17)R_4 = 0.83 \cdot R_4$ in expanding space.

which allows the derivation of the redshift into the forms

$$z = e^\alpha - 1 = \frac{D}{R_{4(2)}} e^\alpha = \frac{D/R_{4(2)}}{1 - D/R_{4(2)}} \tag{47}$$

The maximum optical distance of an object in space is equal to the 4-radius at the time of the observation $D = R_{4(2)}$. Figure 7 illustrates the development of the optical path and the redshift from objects at different $R_4$ radii of space. The optical distance of an object, $D$, can be expressed in terms of the redshift and current 4-radius as

$$D = R_4 e^{-\alpha} = \frac{z}{1+z} R_4 \tag{48}$$

As shown by equation (48), the optical distance of an object approaches the 4-radius $R_4$ at very high redshifts but never exceeds it.

*The recession velocity of distant objects*

The observed recession velocity of distant objects is the optical recession velocity, which is the velocity at which the optical distance increases

$$v_{rec(optical)} = \frac{dD}{dt} = c\left(1 - 1/e^\alpha\right) = \frac{D}{R_4} c \tag{49}$$

As shown by equation (49), the optical recession velocity never exceeds the velocity of light at the time of the observation but approaches it asymptotically.

The physical recession velocity, which means the actual increase of the physical distance at the time of the observation, exceeds the velocity of light for objects at distance angles $\alpha > 1$ radians.

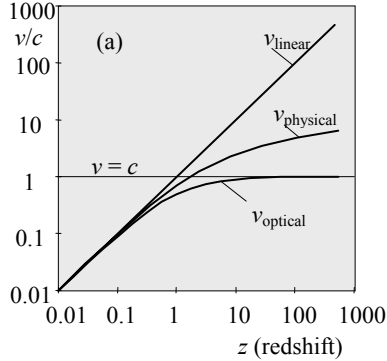

**FIGURE 8.** Recession velocities as a function of redshift. The curves show the recession velocities defined in different ways. The optical velocity, the velocity of the lengthening of the optical distance, never exceeds the velocity of light as it is at the time of the observation but approaches it asymptotically. When the optical recession velocity approaches the velocity of light, the optical distance approaches the length of the 4-radius of space. The physical distance, however, exceeds the velocity of light when the optical distance exceeds 1 radian. Such a situation occurs when the redshift exceeds $z > e^1 - 1 = 1.718$.

$$v_{rec(phys)} = \frac{dD}{dt} = \alpha \frac{dR_4}{dt} = \alpha c \qquad (50)$$

(see Figure 8).

*Observation angle of distant objects*

The optical angle $\theta$ subtended by a standard rod $r_s$ can now be related to the angular size of the rod relative to the center of the 4-sphere as

$$\frac{\theta_{DU(rod)}}{r_s/R_4} = \frac{z+1}{z} \qquad (51)$$

(see FIG. 9). As a major difference to the standard cosmology model, in the dynamic universe the orbital radii of local gravitational systems are subject to the expansion of space. The radii of planetary systems as well as the radii of galaxies expand in direct proportion to the expansion of the 4-radius $R_4$. For example, out of the 3.8 cm annual increase of the Earth to Moon distance about 2.8 cm comes from the expansion of space and only 1 cm from the tidal effects.

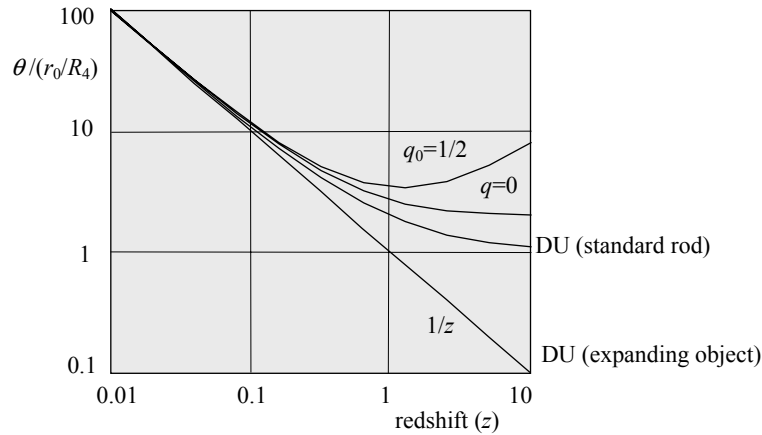

**FIGURE 9.** Normalized observation angle of a standard rod as a function of redshift in the DU model and in the standard model with expansion parameter $q_0$ as $q_0 = \frac{1}{2}$ and $q_0 = 0$. The reference line $1/z$ shows the observation angle of the standard rod in Euclidean space. For objects expanding in space, the prediction for the observation angle obtains the Euclidean $1/z$ form.

The observation angle of expanding objects obtains the Euclidean form

$$\frac{\theta_{DU(exp.obj.)}}{r_{ex}/R_4} = \frac{1}{z} \qquad (52)$$

The prediction of equation (52) is supported by observations of angular sizes of radio sources (see FIG. 10).

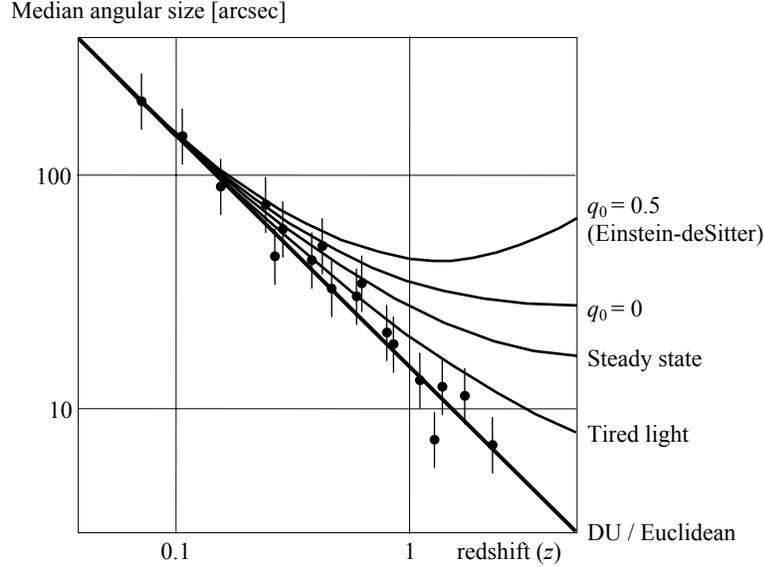

**FIGURE 10.** Comparison of equation (52) with the predictions of the standard cosmology model for various $q_0$ values (without evolution) and the tired light model [7], A. Sandage: The Deep Universe, original data for the median angular sizes (arcsec) and redshifts for radio-sources by Kapahi's (1987). The prediction given by equation (52) is the straight line with Euclidean appearance showing a considerable match with observations.

*The magnitude of standard candles*

Perhaps the most striking recent cosmology observation is the magnitude versus redshift of supernova explosions [8,9,10]. When interpreted with the standard cosmology model, the observations mean that the expansion of space is accelerating instead of decelerating as could be expected due the work done against gravitation. To motivate the acceleration, dark energy in the form of $\Omega_\lambda$ has been added to the expression of the magnitude in the standard model

$$m = M + 5\log\left[\frac{c(1+z)}{H_0}\int_0^z \frac{1}{\sqrt{(1+z)^2(1+\Omega_m z) - z(2+z)\Omega_\lambda}}dz\right] + 25 \qquad (53)$$

Supernovae are regarded as standard candles emitting a fixed number of quanta at any redshift. In spherically closed space the energy density observed from such a source is

$$F_{rad(r)} = \frac{1}{(z+1)^2}\frac{E_{\lambda(e)}}{(z+1)A_e}\left(\frac{z+1}{z}\right)^2 = \frac{F_{rad(e)}}{z^2(z+1)} \qquad (54)$$

where $F_{rad(e)} = E_{\lambda(e)}A_e$ is the energy density of a source of surface area $A_e$. The first two terms in equation (54) describe the dilution of the radiation during propagation and the third term the observed surface as the square of the angular size of a standard rod in equation (51). In terms of the magnitude, equation (54) obtains the form

$$m = M_0 + 5\log(z) + 2.5\log(z+1) \qquad (55)$$

which agrees with observations at least as well as the standard model with optimized $\Omega_m$, $\Omega_\lambda$, and $H_0$. The excellent fit of equation (55) lends strong support to the zero energy balance of closed spherical space (see FIG. 11) [11]. In equation (55), the only parameter is the reference magnitude $m_0$, whereas in the standard cosmology prediction (53) there are, additionally, two density parameters and the Hubble constant as parameters to be optimized.

The fit of equation (55) means that the expansion of space continues at a decelerating rate in a zero energy balance with the diminishing gravitational energy of the structure. Equation (55) does not need any assumptions regarding the mass density in space or the value the Hubble constant.

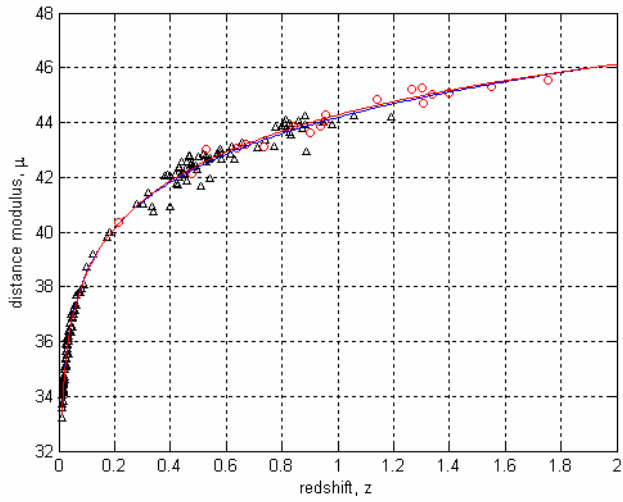

**FIGURE 11.** Distance modulus $\mu = m - M$, vs. redshift for Riess et al's gold dataset and the data from the HST. The triangles represent data obtained via ground-based observations, and the circles represent data obtained by the HST. The optimum fit for the standard cosmology equations is shown in red, and the fit for the equation (55) is shown, slightly below, in blue [11].

*Surface brightness of expanding objects*

In spherically closed dynamic space, galaxies and any local gravitational systems expand with the expansion of space. While maintaining the total luminance, the surface brightness of expanding objects decreases with the expansion of space, i.e., the surface brightness of objects with high redshift is higher than the surface brightness of low redshift objects. By taking into account the Euclidean appearance of the angular surface area of the expanding objects as the square of equation (52), the observed surface brightness of galaxies can be obtained from the received energy density given in equation (54) as

$$B_{S(r)} \sim \frac{F_{rad(e)}}{z^2(z+1)}z^2 \sim \frac{F_{rad(e)}}{z+1} \qquad (56)$$

Equation (56) means that the increase of the actual surface brightness of the object with an increasing redshift is compensated by the Euclidean decrease of the observed angular area of the object, and the only remaining effect is the dilution of the energy density due to the redshift, the $1/(z+1)$ term. The prediction of equation (56) is supported by recent observations of the surface brightness of high redshift galaxies [12].

*Radioactive dating, buildup of large scale structures*

Dilution of the rest energy of matter with the decreasing velocity of light in expanding space means that the rate of all internal atomic processes slows down with the expansion. Also, the rate of radioactive decay decreases with the expansion, which means that the results of radioactive dating must be corrected for a higher decay rate in the past. As demonstrated in Figure 12, the 14 billion year result produced by dating assuming a constant decay rate is reduced to about 9 billion years when taking into account the decreasing decay rate. The reduction solves the currently recognized problem of the age of the oldest stars apparently exceeding the age of expanding space.

The higher energetic state of the early expanding universe means also that all velocities in space have been higher. This is of special importance when estimated the buildup times of the large scale galactic structures in space.

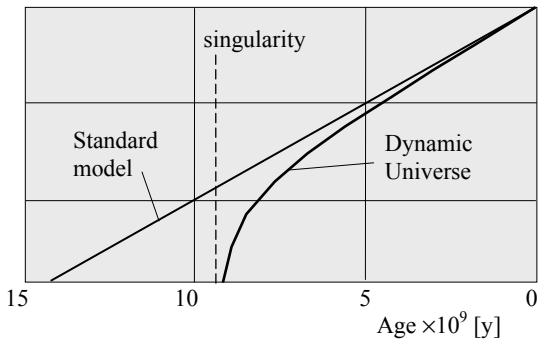

**FIGURE 12.** Accumulation of decay products according to the standard model with a constant decay rate and the DU model with a decreasing decay rate.

*Microwave background radiation*

In spherically closed dynamic space, the background radiation means radiation emitted from an observer's location in space at redshift

$$z = e^{n \cdot 2\pi} - 1 = 535.5^n - 1 \qquad (57)$$

which means 360° propagation through spherically closed space. For $n = 1$ the redshift of background radiation is $z = 534.5$, corresponding to one full cycle of propagation through spherical space.

The 4-radius of space at the time of the emission of the background radiation was $R_{4(0)} = R_4/535.5 \approx 26$ million light years, which occurred about 750 000 years after the singularity (see FIG. 13).

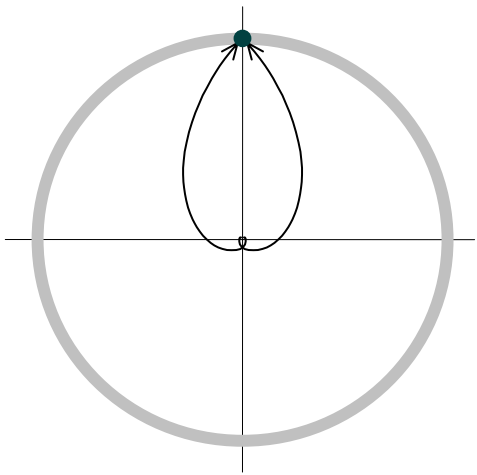

**FIGURE 13.** Background radiation, as propagation of light through a 360° path. The radius of [the] spherically closed space at the time the light was emitted was $R_4 \approx 26$ million light years, and the corresponding time since singularity was $t \approx 750\,000$ years. The velocity of light at that time was about 23 times the present velocity of light.

**CONCLUSIONS**

The Dynamic Universe means major reconsideration of the cosmological structure and development of the universe. It gives a logically consistent, holistic view of observable physical reality. The dynamic universe approach is primarily based on the zero-energy balance of whole space and the conservation of energy in interactions in space. The dynamic universe approach shows relativity in absolute time and distance through the direct effects of gravitation and motion on the locally available rest energy of matter — thus making relativity an integral a part of quantum mechanics.

The Dynamic Universe model describes the local gravitational potential through the geometry of space. The geometry is determined by the conservation of the total gravitational energy. The local curvature of space determines the direction of the local fourth dimension relative to the 4-radius of space. As a consequence, the local curvature of space determines the local imaginary velocity of space and, hence, also the local velocity of light and the locally available rest energy of matter.

Some important conclusions of the physical and cosmological appearance of and phenomena in spherically closed dynamic space can be summarized as follows:

- Universal, absolute time applies to all phenomena in space.
- A local state of rest is a property of a local energy system instead of a property of an inertial observer.
- The rest energy of matter is the energy of motion mass possesses due to the motion of space in the fourth dimension; conservation of the total rest energy in interactions in space relates any state of motion in space to the state of rest in hypothetical homogeneous space.
- The buildup and release of the rest energy of matter can be described as a zero energy process from infinity in the past through singularity to infinity in the future.
- The characteristic emission and absorption frequency due to an electron transition in atomic objects is a function of the velocity and gravitational potential of the atom in the local energy system and the parent systems. As a consequence, coordinate time scales in cascaded gravitational frames and proper times in systems of motion can all be linked to the universal absolute time.
- Electromagnetic resonators can be studied as closed energy systems; as an implication, Michelson–Morley type experiments in moving frames show a zero result.
- Precise predictions for the Shapiro-delay, the perihelion advance of planetary orbits, and the bending of light path near mass centers can be expressed in closed mathematical form.
- The annual Doppler shift of pulsars, the Roemer–effect, and the aberration of starlight get their natural solutions.
- The radii of local gravitational systems expand in direct proportion to the expansion of the 4-radius of space.
- Distant space is observed in Euclidean geometry (e.g. the angular sizes of galaxies).
- The prediction derived for the magnitude versus redshift of a standard emission source gives a perfect fit to recent supernova observations without an assumption of dark energy or indication of accelerating expansion [8].
- The age of expanding space obtains the form $t = 2/3\ R_4/c$.
- Age estimates obtained by radioactive dating are reduced due to the higher decay rate in the young universe (the decay rate is inversely proportional to $t^{1/3}$).

**REFERENCES**

1. Einstein, A., *Kosmologische Betrachtungen zur allgemeinen Relativitätstheorie*, Sitzungsberichte der Preussischen Akad. d. Wissenschaften, 1917
2. Feynman, R., Morinigo,W., Wagner,W., Feynman Lectures on Gravitation (during the academic year 1962-63) , Addison-Wesley Publishing Company, 1995, p. 164
3. Feynman, R., Morinigo,W., Wagner,W., Feynman Lectures on Gravitation (during the academic year 1962-63) , Addison-Wesley Publishing Company, 1995, p. 10
4. Tuomo Suntola, *Theoretical Basis of the Dynamic Universe*, ISBN 952-5502-1-04, 292 pages, Suntola Consulting Ltd., 2004, http://www.sci.fi/~suntola/
5. Tuomo Suntola, "Dynamic space converts relativity into absolute time and distance", Physical Interpretations of Relativity Theory IX (PIRT-IX), London, 3-9 September 2004, http://www.sci.fi/~suntola/DU%20library/PIRT-IX%20reprint.pdf
6. Tuomo Suntola, "Photon – the minimum dose of electromagnetic radiation", Proceedings of SPIE Vol. 5866, 2005, pp. 17-25, http://www.sci.fi/~suntola/DU%20library/Paper%205866-4,%20reprint.pdf

7.  A. Sandage: The Deep Universe, original data for the median angular sizes (arcsec) and redshifts for radio-sources by Kapahi's, 1987
8.  Perlmutter et al., "Measurements of Omega and Lambda from 42 High-Redshift Supernovae", 1999, Astrophys.J., 517, 565 http://arxiv.org/abs/astro-ph/9812133
9.  Knop et al., "New Constraints on Omega_M, Omega_Lambda, and w from an Independent Set of Eleven High-Redshift Supernovae Observed with HST", 2003, Astrophys.J., 598, 102 http://arxiv.org/abs/astro-ph/0309368
10. Riess et al., "Type Ia Supernova Discoveries at z > 1 From the Hubble Space Telescope: Evidence for Past Deceleration and Constraints on Dark Energy Evolution", Astrophys.J., 2004, 607, 665-687 http://arxiv.org/abs/astro-ph/0402512
11. Tuomo Suntola and Robert Day, *Supernova observations fit Einstein-deSitter expansion in 4-sphere*, http://arxiv.org/abs/astro-ph/0412701
12. R.J. Bouwens et.al., http://arxiv.org/abs/astro-ph/0406562